\documentclass[conference]{IEEEtran}

\usepackage{xcolor}
\usepackage{amsmath}
\usepackage{amsfonts}
\usepackage{booktabs}
\usepackage{makecell}
\usepackage{array}
\usepackage{multirow}
\usepackage{graphicx}
\usepackage{mycommands}
\usepackage{forest}
\usepackage{comment}
\usepackage{xspace}
\usepackage{float}
\usepackage[hidelinks]{hyperref}
\usepackage{xurl}
\usepackage{url}

\usepackage[table]{xcolor}

\usepackage{tikz}
\usetikzlibrary{arrows.meta,positioning,shapes,trees}
\tikzset{
  decision/.style = {rectangle, rounded corners, draw, thick, align=center, inner sep=4pt},
  chance/.style   = {circle, draw, thick, align=center, inner sep=1.5pt},
  terminal/.style = {diamond, draw, thick, aspect=2, inner sep=2pt, align=center},
  edge/.style     = {->, >=Stealth, thick},
  labelnode/.style= {font=\small, inner sep=1.5pt}
}

\begin{document}

\title{Orbital Escalation: Modeling Satellite Ransomware Attacks Using Game Theory}

\author{\IEEEauthorblockN{Efrén López-Morales}
	\IEEEauthorblockA{New Mexico State University\\
		elopezm@nmsu.edu}
        }

\IEEEoverridecommandlockouts
\makeatletter\def\@IEEEpubidpullup{6.5\baselineskip}\makeatother
\IEEEpubid{\parbox{\columnwidth}{
        {\fontsize{7.5}{7.5}\selectfont 
        Workshop on Security of Space and Satellite Systems (SpaceSec) 2026\\
        23 February 2026, San Diego, CA, USA\\
        ISBN 978-1-970672-02-2\\
        https://dx.doi.org/10.14722/spacesec.2026.230006\\
        www.ndss-symposium.org}
}
\hspace{\columnsep}\makebox[\columnwidth]{}}

\maketitle

\begin{abstract}
Ransomware has yet to reach orbit, but the conditions for such an attack already exist.  
This paper presents the first game-theoretic framework for modeling ransomware against satellites: the \emph{orbital escalation game}.  
In this model, the attacker escalates ransom demands across orbital passes, while the defender chooses their best strategy, e.g., attempt a restore procedure.
Using dynamic programming, we solve the defender’s optimal strategy and the attacker’s expected payoff under real orbital constraints. Additionally, we provide a GPS III satellite case study that demonstrates how our orbital escalation game can be applied in the context of a fictional but feasible ransomware attack to derive the best strategies at every step. 
In conclusion, this foundational model offers satellite owners, policy makers and researchers, a formal framework to better prepare their responses when a spacecraft is held for ransom.
\end{abstract}

\IEEEpeerreviewmaketitle

\section{Introduction}

Ransomware remains one of the most disruptive forms of cyberattacks, leveraging control over critical assets to extort payments (ransoms) from victims. The damage caused by ransomware includes loss of data, and business downtime. In 2024 the average ransom payment was \$2.73 million~\cite{sophos2024ransomware}.

While traditionally associated with terrestrial systems such as enterprise servers, hospitals, and critical infrastructure, recent incidents have drawn attention to the growing vulnerability of space-based systems.
The 2022 cyberattack on Viasat's KA-SAT network, which disrupted satellite services across Europe, exemplifies the real-world risk of targeting satellite infrastructure via ground-side vectors \cite{viasat_attack}. 
Additionally, research has shown that ransomware techniques against Low Earth Orbit (LEO) satellites are feasible using vulnerabilities in components like NASA's Core Flight System (cFS)~\cite{donchev2024ransomware,falco2023wannafly}.

A successful ransomware attack on an in-orbit satellite could potentially result in millions of dollars worth of damage. A single Medium Earth Orbit (MEO) satellite may cost hundreds of millions of dollars, serve irreplaceable national functions, and take years to design, launch, and commission. For example in 2022 it was reported that two new GPS III satellites had been ordered for \$744 million~\cite{spacenews_gps744}.
These characteristics make satellite systems attractive to ransomware attackers seeking maximum leverage, and presenting a constrained decision space for satellite owners imposed by short communication windows (orbital passes).

Despite these stratospheric risks, there is no public space policy or guidance on how satellite owners should or even could respond when faced with a ransomware attack~\cite{reese2025space}.

In comparison, such theories have  successfully been developed for terrestrial systems, including, \emph{game theory} models~\cite{li2021game,caporusso2019game}  which provide strategies that help decision-makers respond when faced with a ransomware attack. 

Game theory is a mathematical framework for modeling interactions between rational decision-makers~\cite{von1944theory}. It provides tools to analyze how agents with potentially conflicting objectives choose actions, anticipate responses, and optimize outcomes. In the context of computer security, game theory allows us to systematically study attacker-defender dynamics. However, to the best of our knowledge, game theory has not been applied to satellite ransomware attacks.

To solve this problem, this paper presents a game-theoretic model for satellite ransomware attacks that formalizes the interaction between the attacker, i.e., ransomware adversary, and defender, i.e., satellite owner. 
Our proposed model is designed as a Stackelberg game~\cite{osborne2004introduction} (a sequential leader-follower game) where a particular satellite's communication window or orbital pass is used as the basis for attacker-defender strategies. 
At each orbital pass the attacker escalates their ransomware demands, i.e., increases the ransom and the defender chooses a strategy, such as a recovery procedure. 

In addition to encoding LEO and MEO satellites' communication constraints, our model integrates 7 economic factors such as the downtime cost while the hostage satellite is not operational, and operational constraints such as limited restore procedures.
By combining these temporal and strategic dependencies, our model captures how ransomware attacks can evolve across orbital passes, and how defenders can optimally time recovery actions to minimize total expected loss.
We call this model, the \emph{orbital escalation game}.

To evaluate our model, we develop a case study where a fictional Advanced Persistent Threat (APT) successfully takes a GPS III satellite hostage and use our model to explore multiple attack outcomes. Our case study shows how our model could be applied to potential ransomware incidents.

In summary, this paper makes the following contributions:
\begin{itemize}

    \item We formulate, to the best of our knowledge, the first game-theoretic model for satellite ransomware attacks.

    \item We solve our model's equilibrium, i.e., the optimal strategies for attacker and defender, and discuss how these strategies can inform operational decisions.

    \item We provide a case study, that illustrates how our model can be applied to real-world satellite ransomware attacks. 
    
    \item Finally, we provide an open-source analysis tool that implements our model\footnote{https://github.com/efrenlopezm/orbital-escalation-game}. 
    
\end{itemize}

\section{Background}

\subsection{Ransomware Threats to Satellites}\label{background:ransomwareinsatellites}

While ransomware is well studied in terrestrial systems~\cite{caporusso2019game,li2021game,laszka2017economics,liu2025teamwork}, recent work has started to explore how these attacks unfold in the context of satellites.

Falco et al.~\cite{falco2023wannafly} introduced \emph{WannaFly}, a proof-of-concept ransomware attack tailored for LEO satellites. Their work demonstrated how a malicious payload uploaded to a satellite could mimic the logic of terrestrial ransomware by denying access to mission services until a ransom is paid. While the attack did not rely on strong encryption of the entire system, it effectively disrupted the availability of satellite resources.

Donchev et al.~\cite{donchev2024ransomware} extended this threat model by describing scenarios where ransomware targets satellite ground and space segments simultaneously. They emphasized how attackers might manipulate telecommands (TCs), to impose ransom demands, creating a multi-vector threat surface. 

Together, these studies suggest that ransomware in satellites differs fundamentally from terrestrial settings. First, satellites often lack native full-disk or system-wide encryption, preventing attackers from using encryption-based ransomware. For example, the NanoMind A3200 on-board computer (OBC), widely used in CubeSats, includes an ARM9 processor~\cite{gomspaceA3200} which does not support native encryption.
Second, this limitation allows for alternative recovery strategies. For example, in 2024 a researcher managed to restore communications with a satellite that was considered lost~\cite{beesatRecovery}. Rather than requiring a decryption key, satellite operators could leverage safe-mode functions, and redundancy systems~\cite{eoportalGioveA}. These restore functionalities are essential when modeling satellite ransomware attacks. The differences between terrestrial and satellite ransomware attacks are depicted in Table~\ref{tab:ransomware_comparison}.

\subsection{Game Theory Model Building Blocks}\label{background:buildingblocks}

Game theory uses fundamental components that define the interactions between players which we now describe.

\textbf{Players.} Players are the decision-makers in the game. They can be individuals, nation-states, or satellite operators~\cite{osborne2004introduction}. In the context of ransomware attack models, players are labeled as ``attacker'' and ``defender.'' The \emph{attacker} launches the ransomware attack, and the \emph{defender}, e.g., the satellite operator, reacts to the attacker's actions and makes strategic decisions to recover the satellite held for ransom. 

\textbf{Strategies.} Strategies are the available actions each player can choose from at a particular stage in the game. Strategies are further categorized between \textit{pure strategies} and \textit{mixed strategies}~\cite{osborne2004introduction}. Pure strategies are deterministic and their outcome is completely predictable. For example the ``pay ransom'' strategy always ends the game. Conversely, mixes strategies are nondeterministic and depend on probability. For example, the ``recovery procedure'' strategy may or may not succeed based on some previously known probability, e.g., 0.7.

\textbf{Payoffs and Expected Costs.}  Payoffs quantify how desirable an outcome is to a player and players select strategies to maximize their payoffs, e.g., ransom payment~\cite{osborne2004introduction}. Although payoffs apply to all players in the game, in cost-oriented settings, such as a ransomware attacks, it is more intuitive to model the defender as minimizing an \emph{expected cost}, which plays a role equivalent to a payoff.

For example, if two players choose strategies that lead to a particular outcome, one player’s payoff might be positive (a gain), while the other’s is negative (a loss).  
In this paper, the attacker will be modeled as maximizing \emph{payoff}, while the defender will be modeled as minimizing \emph{expected cost}.

\textbf{Information Set.} The information set determines what each player knows at different stages of the game. Game theory models may have different assumptions about player information. For example, Stackelberg games, by definition assume that all players have \textit{perfect} information, where each player knows all previous moves by all players at every decision point~\cite{osborne2004introduction}. Because our model is a Stackelberg game, both attacker and defender have perfect information.

\textbf{Equilibrium.} The equilibrium is the optimal solution to the game assuming rational players. One game may have multiple equilibria, and each equilibrium differs depending on timing, information, and assumptions. Equilibria must be calculated which is also commonly known as \emph{solving the game}. This involves finding a set of strategies where no player can improve their outcome by unilaterally deviating~\cite{osborne2004introduction}.

\subsection{Stackelberg Games}\label{background:typesofgames}

Game theory has multiple types of games such as Nash games~\cite{nash1950equilibrium}, where all players implement their strategies simultaneously and are often symmetric, e.g., all players use the same strategies. On the other hand, Stackelberg games, are sequential, e.g.,  players observe the opponent's strategy and then react, and they are asymmetric, e.g., players have different strategies. We model our orbital escalation game as a Stackelberg because ransomware interactions are inherently sequential and asymmetric, as such this type of games provide a natural framework to build our model.

Formally, in Stackelberg games, both players are fully rational and the follower has perfect information about the leader's move~\cite{stackelberg1934marktform}. The leader, anticipating the follower's best response, chooses a strategy that maximizes their own utility given this knowledge. The solution to a Stackelberg game is the subgame perfect Nash equilibrium (SPNE)~\cite{fudenberg1991game}. 
Lastly, in Stackelberg games, the concept of a \textit{horizon} is used to define how far into the future the game extends. A finite horizon limits the number of sequences that can occur (e.g., the number of orbital passes).
\section{Threat Model}

In our orbital escalation game, we consider an attacker capable of launching ransomware attacks against LEO or MEO satellites (e.g. state-sponsored actors) with the objective of coercing the satellite operator into meeting specific demands, e.g., ransom payment.
The attacker may exploit vulnerabilities in the satellite's command and control infrastructure, ground segment interfaces, or on-board software.
While terrestrial ransomware attacks almost always involve encrypting the system's data we consider any attack in which an adversary blocks the legitimate satellite owner from accessing and operating the spacecraft, which we discussed in Sec.~\ref{background:ransomwareinsatellites}. 

We also assume that the attacker has knowledge of the location(s) of the satellite's mission ground stations and the satellite's orbit which allows the attacker to predict future passes and their duration from both the attacker's location and the defender's ground stations.

For the defender or satellite operator, we consider that they maintain the capability to attempt to communicate with the spacecraft via an operational mission control system.

\section{Orbital Escalation Game: Model Formulation}\label{sec:model}

We model our orbital escalation game as a Stackelberg game between the ransomware attacker and the satellite operator defender. The key feature of our model is that both the attacker and defender strategies revolve around the hostage satellite's orbit. The attacker \emph{escalates} their ransom every time a pass is completed and the defender can only implement a strategy during orbital passes, when ground station communication is possible.
We now formally describe our model.

\subsection{Players and Strategies}\label{subsec:playersandstrategies}

There are two players, the ransomware attacker and the satellite operator defender.
The attacker initiates the game and commits up front to an extortion strategy defined as $\pi_A = (R_0, \Delta R)$, where $R_0$ is the initial ransom demand and $\Delta R$ is the increment added to the ransom after each orbital pass. Additionally, the attacker pays a one-time exploit cost $C_{\mathrm{atk}}$ to compromise the satellite and incurs a holding cost $c_{\mathrm{hold}}$ per orbital pass while the game continues. This holding cost may be due to the attacker's operational costs, e.g., electricity, however, it may also increase due the risks associated with their illegal activity, (the cost of getting caught).

At each orbital pass, the defender must select a strategy to respond. The available defender strategies are:
    \begin{enumerate}
        \item \textsc{Pay}: immediately transfer the ransom $R_k$ to the attacker. This strategy ends the game.
        \item \textsc{Idle}: take no direct action, instead use the pass for indirection action, e.g., plan or communicate internally, incurring a downtime cost $c_{\downarrow}$ until the next orbital pass.
        \item \textsc{Restore $j$}: attempt a recovery procedure $j$ that has a direct cost $C_j$, requires $d_j$ passes to complete, and succeeds with probability $p_j$.
        \item \textsc{Refuse}: declare refusal to pay. This terminates the game and incurs a mission loss $L_{\mathrm{ref}}$ for the defender.
    \end{enumerate}

\subsection{Ransom Escalation Dynamic}

The game's time dynamic or cadence is controlled by orbital passes, denoted $\{t_k\}_{k=1}^K$ where $K$ defines the \emph{horizon} of the game (maximum number of passes) as discussed in Sec~\ref{background:typesofgames}. At each pass $k$, the attacker’s ransom has escalated to:

\begin{figure}[H]
\centering
\[
R_k = R_0 + (k-1)\Delta R
\]
\caption{Current ransom at orbital pass $k$. }
\label{fig:current-ransom}
\end{figure}

Thus, waiting without paying is costly, both because downtime cost accumulates for the defender, and because ransom steadily increases. A shorter horizon compresses decision-making, forcing earlier payments or risk of mission loss, while a longer horizon creates opportunities to attempt recoveries.

\subsection{Equilibrium Analysis}

As with any game theory model, we need to solve our orbital escalation game's equilibrium,  discussed in Sec.~\ref{background:buildingblocks}. 
To achieve this, we need to analyze it from the perspectives of both attacker and defender.

First, we need to solve the \emph{defender’s decision problem} which yields the \emph{expected cost} (Sec.~\ref{background:buildingblocks}) and the corresponding optimal strategy. 
Second, given the optimal strategy, we need to solve for the the attacker’s \emph{expected payoff}.

\subsubsection{Defender’s Expected Cost}\label{subsec:defender-problem}

Solving the defender's decision problem provides the best-response strategy under any given attacker policy by minimizing the defender's expected cost, and it is necessary to solve the game equilibrium.

To solve the defender's decision problem we use dynamic programming. Specifically, we use the Bellman equation~\cite{bellman2015applied}, which is a standard technique in dynamic programming that breaks a problem into a sequence of simpler subproblems.

We model the defender's decision problem as the Bellman recursive equation shown in Fig.~\ref{fig:bellman}. Let $V_k$ represent the defender’s \emph{minimal expected cost} starting from pass $k$. At each decision point, the operator weighs four actions' costs: \textsc{Pay}, \textsc{Idle}, \textsc{Restore} and \textsc{Refuse} as described in Sec.~\ref{subsec:playersandstrategies}.

\begin{figure*}[h]
\centering
\[
V_k = \min\Bigg\{
\underbrace{R_k}_{\textsc{Pay}}, \;\;
\underbrace{c_{\downarrow} + V_{k+1}}_{\textsc{Idle}}, \;\;
\underbrace{\min_{j}\Big[C_j + d_j c_{\downarrow} + (1-p_j)V_{k+d_j}\Big]}_{\textsc{Restore}}, \;\;
\underbrace{L_{\mathrm{ref}}}_{\textsc{Refuse}}
\Bigg\}.
\]

\caption{Defender’s decision problem equation modeled as a Bellman recursion at orbital pass $k$. 
Here $V_k$ is the defender’s minimal expected cost; 
$R_k = R_0 + (k-1)\Delta R$ is the ransom demand at pass $k$; 
$c_{\downarrow}$ is the per-pass downtime cost; 
$V_{k+1}$ and $V_{k+d_j}$ are continuation values after idling or a failed restore, respectively; 
$C_j$, $d_j$, and $p_j$ are the direct cost, duration (in passes), and success probability of restore strategy $j$; 
and $L_{\mathrm{ref}}$ is the mission loss if the defender refuses to pay. 
At each pass the defender compares the four options and chooses the action that minimizes cost.}
\label{fig:bellman}
\end{figure*}

If the game has not ended and the end of the horizon ($K+1$) is reached, the defender is forced to simply compare between paying the escalated ransom or suffering mission loss as described in Fig.~\ref{fig:end-horizon}.

\begin{figure}[H]
\centering
\[
V_{K+1} = \min\{L_{\mathrm{ref}}, R_{K+1}\}
\]
\caption{Defender's end of horizon decision. }
\label{fig:end-horizon}
\end{figure}

Together Eq.~\ref{fig:bellman} and Eq.~\ref{fig:end-horizon} capture the defender’s best-response policy $\mu^*$: a mapping from each orbital pass to the least costly available action.

\subsubsection{Attacker’s Expected Payoff}

Given the defender’s optimal response policy $\mu^*$, the attacker’s objective is to 
maximize their expected payoff by selecting ransom parameters 
$\pi_A = (R_0, \Delta R)$. For example, in prior ransomware models~\cite{laszka2017economics}, a rational attacker chooses an initial ransom below the defender’s refusal loss to make payment the defender's preferred response, and setting a ransom escalation rate higher than the downtime costs renders delayed payment unattractive.

Let $A_k$ denote the attacker’s \emph{expected continuation payoff} at orbital pass $k$, 
that is, the expected profit from this pass onward, excluding any past (sunk) costs. 
The attacker pays a one-time exploit cost $C_{\mathrm{atk}}$ when compromising the satellite 
and a holding cost $c_{\mathrm{hold}}$ per orbital pass while the satellite remains under 
their control.  Conditioned on the defender’s chosen action at pass $k$, the attacker’s continuation value 
evolves as follows:

\begin{figure}[h]
\centering
\[
A_{K+1} =
\begin{cases}
R_{K+1}, & \text{if } R_{K+1} \le L_{\mathrm{ref}},\\[4pt]
0, & \text{otherwise.}
\end{cases}
\]

\caption{Attacker's end of horizon payoff.}
\label{fig:horizonattackerpayoff}
\end{figure}

\begin{figure}
\centering
\[
A_k =
\begin{cases}
R_k, & \text{if \textsc{Pay}},\\[4pt]
-\,c_{\mathrm{hold}} + A_{k+1}, & \text{if \textsc{Idle}},\\[4pt]
-\,c_{\mathrm{hold}}\,d_j + (1-p_j)A_{k+d_j}, & \text{if \textsc{Restore }j},\\[4pt]
0, & \text{if \textsc{Refuse}}.
\end{cases}
\]
\caption{Recursive formulation of the attacker’s continuation value $A_k$ at orbital pass $k$. Each term represents the attacker’s expected payoff given the defender’s chosen action (Pay, Idle, Restore, or Refuse).}
\label{eq:attacker-continuation}
\end{figure}

At the end of the horizon, the attacker’s payoff depends on whether the defender 
ultimately decides to pay the ransom or to accept mission loss. 
This terminal condition is defined in Eq.~\eqref{fig:horizonattackerpayoff}. The attacker’s total expected payoff from the start of the game is calculated using Eq.\eqref{fig:total-payoff}.

\begin{figure}[H]
\centering
\[
\Pi_{\mathrm{att}} = -C_{\mathrm{atk}} + A_{k_0}
\]
\caption{The attacker’s total expected payoff $\Pi_{\mathrm{att}}$, combining the initial exploit cost $C_{\mathrm{atk}}$ and continuation value $A_{k_0}$ at the first pass.}
\label{fig:total-payoff}
\end{figure}

\subsection{Equilibrium Solution}\label{subsec:equilibrium-solution}

The game reaches an \emph{equilibrium} when both players’
strategies are mutually optimal given the sequential order of play. 

The attacker first commits to a ransom policy 
$\pi_A = (R_0, \Delta R)$, where $R_0 \ge 0$ and $\Delta R \ge 0$.
Given this policy, the defender observes $\pi_A$ and plays a best-response 
policy $\mu^*$. The attacker then anticipates this response and chooses the ransom policy that 
maximizes their total expected payoff:

\begin{figure*}[h]
\centering
\[
\pi_A^\star = 
\underbrace{\arg\max_{R_0,\,\Delta R \ge 0}}_{\text{Optimization of ransom policy}} 
\Bigg\{
\underbrace{-\,C_{\mathrm{atk}}}_{\text{Exploit cost}}
+\;
\underbrace{A_{k_0}(\pi_A, \mu^*)}_{\text{Expected payoff given defender's best response}}
\Bigg\}.
\]

\caption{
Attacker’s optimization problem equation. 
Here $\pi_A = (R_0, \Delta R)$ is the attacker’s ransom policy, where 
$R_0$ is the initial ransom demand and $\Delta R$ is the increment applied after each orbital pass. 
$C_{\mathrm{atk}}$ represents the one-time exploit cost paid to compromise the satellite, and 
$A_{k_0}(\pi_A, \mu^*)$ is the attacker’s expected continuation payoff starting from the first 
orbital pass $k_0$, given the defender’s optimal response policy $\mu^*$. 
The symbol $\arg\max$ indicates that the attacker chooses the specific values of $(R_0, \Delta R)$ 
that maximize the total expected payoff.
}

\label{fig:attacker-optimization}
\end{figure*}

At equilibrium, the pair $(\pi_A^\star, \mu^*)$ satisfies:
\begin{itemize}
    \item The defender’s strategy $\mu^*$ is the optimal (lowest-cost) response 
    to the attacker’s ransom policy $\pi_A^\star$.
    \item The attacker’s strategy $\pi_A^\star$ yields the highest expected payoff 
    given that the defender responds optimally.
\end{itemize}

\subsection{Model Assumptions}

To make the orbital escalation game analytically tractable while preserving the key features of a realistic satellite ransomware scenario, we adopt the following four assumptions. 
First, the game is discrete and sequential in time, unfolding across orbital passes. Second, the model assumes complete information: both players know all relevant parameters of the game at all times, which is a necessary feature of Stackelberg games (Sec.~\ref{background:typesofgames}). Although in an operational setting the players will not actually have access to perfect information, our model serves as a conservative worst-case baseline. Third, costs and rewards are additive across passes. Fourth and final, the game terminates when the defender chooses \textsc{Pay}, \textsc{Refuse}, when a 
restoration attempt succeeds or when the maximum horizon $K$ is reached without resolution.

Our model's multiple parameters are summarized in Appendix~\ref{appendix:parameterstable} for better readability.

\section{Case Study: \examplesat Ransomware Attack}
\label{sec:worked-example-once}

In order to illustrate how our orbital escalation game works, we instantiate it via a case study where a real-world satellite, a \examplesat, becomes the target of a fictional ransomware attack.
We selected the \examplesat because there is enough public documentation available regarding its cost and contribution to the US economy which allows us to create a more realistic case study. Due to space limitations we are unable to explain the reasoning for each of the parameters' values here, however, they are available in Appendix~\ref{appendix:case-study-parameters}.

\subsection{Parameters}

In this case study the defender is the operator of the GPS III constellation, namely, the U.S. Space Force, and the parameters are as follows:
\newpage
\begin{itemize}
    \item Mission loss ($L_{\mathrm{ref}}) = \text{\$450M}$
    \item Downtime $c_{\downarrow} = \text{\$3M/pass}$
    \item Restore options (each usable once):
    \begin{enumerate}
        \item Safe mode: $C_1 = \text{\$10k}$, $d_1=2$, $p_1=0.9$.
        \item Privileged TC: $C_2 = \text{\$10k}$, $d_2=1$, $p_2=0.4$.
    \end{enumerate}
\end{itemize}

On the other side, the attacker is \attacker, a fictional state-sponsored Advanced Persistent Threat (APT) capable of successfully launching a ransomware attack against a \examplesat. The attacker has the following parameters:

\begin{itemize}
    \item Attacker ransom policy: $R_0 = \text{\$112.5M}$, $\Delta R = \text{\$6M}$.
    \item Attacker costs: $C_{\mathrm{atk}} = \text{\$14M}$, $c_{\mathrm{hold}} = \text{\$10k/pass}$.
\end{itemize}

\attacker also informs the U.S. Space Force they have only 4 passes to pay the ransom otherwise the \examplesat will be permanently lost. This sets the game horizon to ($K=4$), where each restore strategy can be used only \emph{once}. 

\subsection{Solving the Defender's Optimal Response}

Using the previously described parameters, we now solve the orbital escalation game. 

\subsubsection{Backward Induction Procedure}\label{subsec:backward-induction}

To obtain the defender’s optimal strategy, we solve the Bellman Eq.~\eqref{fig:bellman} by \emph{backward induction}, starting from the last orbital pass
and moving toward the first. This means we first determine what the defender would optimally do
if the game reached the end of the horizon, and then propagate those values backward so that every
earlier decision accounts for the future consequences.

\textbf{Terminal condition (K = 5).} At the end of the four-pass horizon ($K=4+1$), if no decision has terminated the game,
the defender faces one final comparison between paying the ransom or suffering total mission loss. The terminal condition value is calculated using Eq.~\eqref{fig:end-horizon}. To calculate the ransom at the terminal orbital pass, we use Eq.~\eqref{fig:current-ransom} and substitute the values of \attacker's ransom policy:
\[
R_5 = R_0 + (5-1)\Delta R = \$112.5\,\text{M} + (5-1)\times\$6\,\text{M} = \$136.5\,\text{M}
\]

Since $R_5 < L_{\mathrm{ref}}=\$450\,\text{M}$, the minimal cost at the terminal step is
$V_5=\$136.5\,\text{M}$. This means that if the defender reaches the end without restoring the satellite,
paying the ransom at that point is less costly than losing the mission outright.

\textbf{Final Pass (K = 4).} Working backward, we evaluate each preceding orbital pass, in this case, $k=4$, by applying the defender's decision problem Eq.~\eqref{fig:bellman}.
At every pass, the defender compares four actions: \textsc{Pay}, \textsc{Idle}, \textsc{Restore} and \textsc{Refuse} as discussed in Sec.~\ref{subsec:playersandstrategies}.

At pass $k=4$ only one communication window remains before the game terminates.
The defender compares:
\begin{center}
\small
\begin{tabular}{@{}lll@{}}
\toprule
\textbf{Defender's Strategy} & \textbf{Formula from Eq.\eqref{fig:bellman}} & \textbf{Cost (\$M)} \\ 
\midrule
\textsc{Pay} &
\begin{tabular}[t]{@{}l@{}}
$R_4 = R_0 + (4-1)\Delta R$ \\ 
$= 112.5 + (3\times6)$
\end{tabular} &
$\$130.5\,\text{M}$ \\
\midrule

\textsc{Idle} &
\begin{tabular}[t]{@{}l@{}}
$c_{\downarrow}+V_5$ \\ 
$= 3+136.5$
\end{tabular} &
$\$139.5\,\text{M}$ \\
\midrule

\textsc{Restore 1 (SM)} &
\begin{tabular}[t]{@{}l@{}}
$d_1=2>1$ \\ 
(not enough passes)
\end{tabular} &
N/A \\
\midrule

\rowcolor{green!10}
\textsc{Restore 2 (PT)} &
\begin{tabular}[t]{@{}l@{}}
$C_2 + d_2 c_{\downarrow}$ \\ 
$+ (1-p_2)V_{4+d_2}$ \\ 
$= 0.01 + 3 + (0.6\times136.5)$
\end{tabular} &
$\$85.91\,\text{M}$ \\
\midrule

\textsc{Refuse} &
\begin{tabular}[t]{@{}l@{}}
$L_{\mathrm{ref}} = 450$
\end{tabular} &
$\$450\,\text{M}$ \\
\bottomrule
\end{tabular}
\normalsize
\end{center}

The minimal expected cost is therefore $V_4 = \$85.91\,\text{M}$, achieved by
\textsc{Restore 2 (PT)}.
Operationally, this represents a last-minute attempt to recover the satellite with the
quick one-pass privileged TC before the final ransom deadline.

\textbf{Pass (K = 3).} At this stage, both restore options remain available since neither has been used yet. 
Safe Mode ($d_1=2$) and Privileged TC ($d_2=1$) are both feasible because they can complete before the terminal step ($k+d_j \le K+1 = 5$). 
The continuation value from the next pass is $V_4 = \$85.91\,\text{M}$. At pass $k=3$ defender compares:

\begin{center}
\small
\setlength{\tabcolsep}{5pt}
\begin{tabular}{@{}lll@{}}
\toprule
\textbf{Defender's Strategy} & \textbf{Formula from Eq.\eqref{fig:bellman}} & \textbf{Cost (\$M)} \\
\midrule
\textsc{Pay} &
\begin{tabular}[t]{@{}l@{}}
$R_3 = R_0 + (3-1)\Delta R$ \\ 
$= 112.5 + (2\times6)$
\end{tabular} &
$\$124.5\,\text{M}$ \\
\midrule
\textsc{Idle} &
$c_{\downarrow}+V_4 = 3+85.91$ &
$\$88.91\,\text{M}$ \\
\midrule
\rowcolor{green!10}
\textsc{Restore 1 (SM)} &
\begin{tabular}[t]{@{}l@{}}
$C_1 + d_1 c_{\downarrow} + (1-p_1)V_{3+d_1}$\\
$= 0.01 + 6 + (0.1\times136.5)$
\end{tabular} &
$\$19.66\,\text{M}$ \\
\midrule
\textsc{Restore 2 (PT)} &
\begin{tabular}[t]{@{}l@{}}
$C_2 + d_2 c_{\downarrow} + (1-p_2)V_{3+d_2}$\\
$= 0.01 + 3 + (0.6\times85.91)$
\end{tabular} &
$\$54.56\,\text{M}$ \\
\midrule
\textsc{Refuse} &
$L_{\mathrm{ref}} = 450$ &
$\$450\,\text{M}$ \\
\bottomrule
\end{tabular}
\normalsize
\end{center}

The minimal expected cost is therefore $V_3 = \$19.66\,\text{M}$, achieved by \textsc{Restore 1 (SM)}. 
Operationally, this means that at the third orbital pass, the defender initiates the longer but highly reliable Safe Mode recovery, which completes just before the final ransom deadline. 
Its high success probability ($p_1 = 0.9$) and low direct cost make it preferable to both paying and the riskier one-pass Privileged TC.

\textbf{Remaining Passes (K = 2 and 1).} Due to space limitations we are unable to develop the remaining two passes. Nevertheless, the remaining passes are calculated using the same method and the summarized results are shown in Table~\ref{tab:pass-summary}.

\begin{table}[t]
\centering
\caption{Backward induction solution for each orbital pass}
\small
\setlength{\tabcolsep}{3pt} 
\begin{tabular}{@{}c p{1.3cm} p{2.2cm} p{1.8cm} c@{}}
\toprule
\textbf{Pass} &
\textbf{Avail. Restores} &
\textbf{Feasible Actions} &
\textbf{Optimal Strategy} &
\textbf{$V_k$ (\$M)} \\
\midrule
5 (term.) & -- & Pay, Refuse &
\textsc{Pay} & $\$136.5\,\text{M}$ \\
\midrule
4 & \{SM,PT\} & Pay, Idle, PT, Refuse &
\textsc{Restore 2 (PT)} & $\$85.91\,\text{M}$ \\
\midrule
3 & \{SM,PT\} & Pay, Idle, SM, PT, Refuse &
\textsc{Restore 1 (SM)} & $\$19.66\,\text{M}$ \\
\midrule
2 & \{SM,PT\} & Pay, Idle, SM, PT, Refuse &
\textsc{Restore 1 (SM)} & $\$14.6\,\text{M}$ \\
\midrule
1 & \{SM,PT\} & Pay, Idle, SM, PT, Refuse &
\textsc{Restore 1 (SM)} & $\$13.9\,\text{M}$ \\
\bottomrule
\end{tabular}

\label{tab:pass-summary}
\end{table}

\subsection{Calculating the Attacker's Payoff}

Under the U.S. Space Force's optimal strategy derived above, the defender never chooses 
to pay the ransom and instead relies on restoration strategies. Consequently, \attacker receives no positive ransom income, yet still incurs the exploit and holding costs. Using Eq.~\eqref{fig:total-payoff} and the recursive formulation in Eq.~\eqref{eq:attacker-continuation}. we calculate the attacker's payoff by substituting the parameters $c_{\mathrm{hold}} = \$10{,}000 = \$0.01\,\text{M}$ (holding cost), $C_{\mathrm{atk}} = \$14\,\text{M}$ (exploit cost) and $K = 4$ (number of passes):

\[
\Pi_{\mathrm{att}} = -C_{\mathrm{atk}} + A_{k_0}
  = -14 - 0.04
  = -\$14.04\,\text{M}
\]

Therefore, \attacker ends the game with a \emph{negative expected payoff} of $-\$14.04\,\text{M}$ due to the fact that the U.S. Space Force's optimal strategy does not include \textsc{Pay} or \textsc{Refuse}. 

\subsection{GPS III Satellite Case Study Summary}

Overall, the case study equilibrium reveals that the U.S. Space Force should prioritize restoration attempts rather than payment with Safe Mode being the optimal strategy 
during the first three orbital passes, while Privileged Telecommand being optimal during the final pass.

However, these results represent the ideal equilibrium strategies assuming that the restore attempts are successful. In a real operational setting, the outcome of the game could change due to factors such as failed restore attempts, or communication delays. We explore these in the next section.

This case study is an illustrative application of our model rather than a general claim about ransomware response strategies across all space missions. The equilibrium outcome observed for GPS III is shaped by mission-specific assumptions, e.g., mission loss cost. Different mission profiles, such as lower cost satellites, may yield substantially different equilibria.

\section{Realized \examplesat scenarios}
\label{sec:expost}

\begin{figure}
    \centering
    \includegraphics[width=0.9\linewidth]{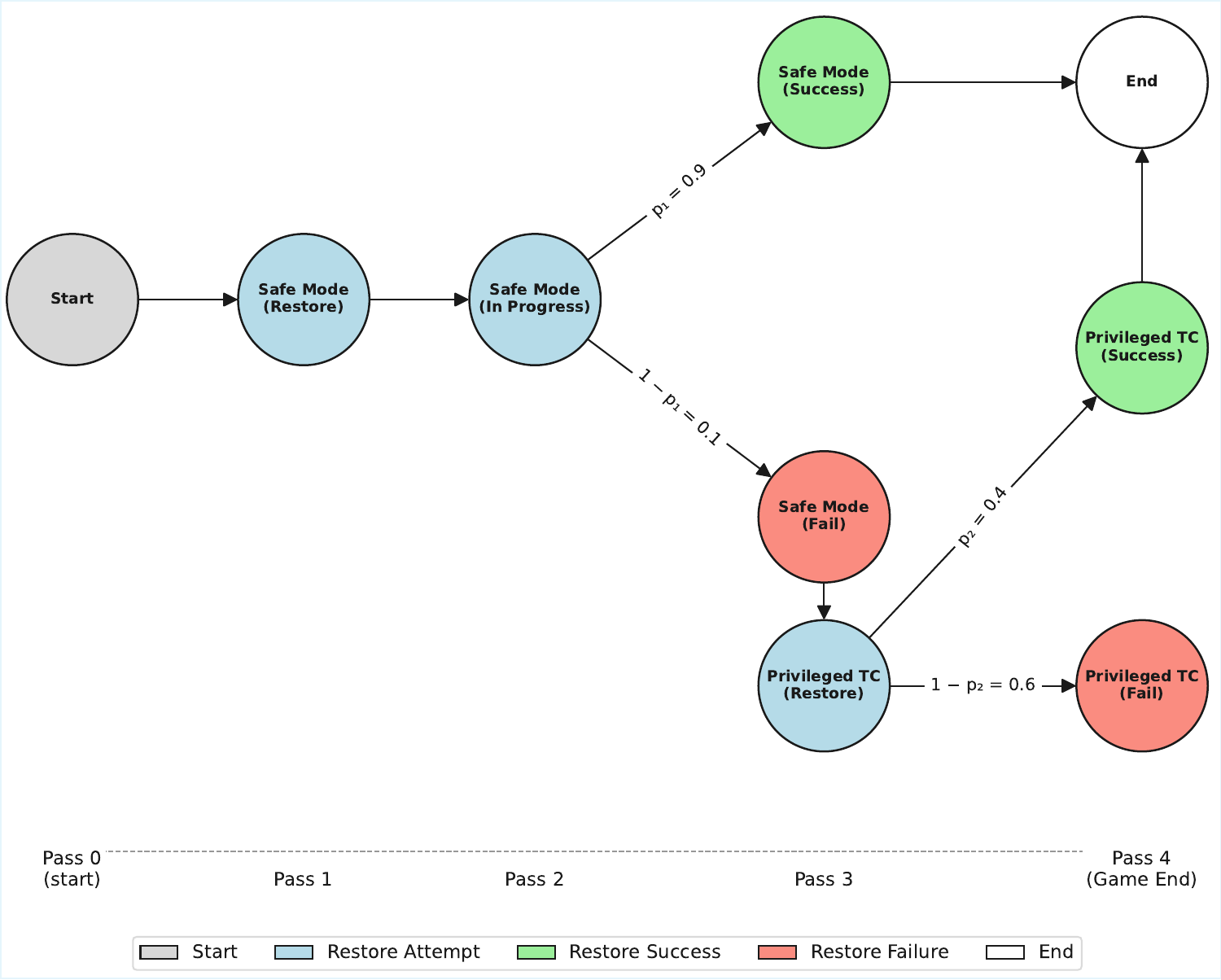}
    \caption{Prepared defender scenario decision tree. The edges that originate from restore attempt nodes include their success probabilities from Sec.~\ref{sec:worked-example-once}.}
    \label{fig:tree}
\end{figure}

In the previous section (Sec.~\ref{sec:worked-example-once}) we obtained the \emph{equilibrium under ideal circumstances}, in  this section we present two scenarios with \emph{realized outcomes} of the \examplesat case study. The purpose is to translate the mathematical equilibrium into an operational narrative that shows how the defender's and attacker's strategies unfold once random outcomes, such as restore success or failure, occur.
Specifically, we present an \emph{unprepared defender} and a \emph{prepared defender} scenario. In the unprepared defender scenario, the U.S. Space Force is not ready to implement their restore strategies, i.e., there is no standard operating procedure, while in the prepared defender scenario, clear procedures are in place to initiate the restore procedures.

\subsection{Unprepared Defender}

We now describe the strategies that the \emph{unprepared defender} (U.S. Space Force) implements during the available four passes. At each pass we compare the \emph{realized selected strategy} vs the strategy recommended by the equilibrium under ideal circumstances as shown in Table~\ref{tab:pass-summary}.

At \textbf{Pass~1}, the equilibrium recommends \textsc{Restore (SM)}, i.e., safe mode. However, the U.S. Space Force is forced to instead implement \textsc{Idle} because they need time to find out how to initiate safe mode. This incurs downtime $c_{\downarrow} = \$3$\,M, for the defender and a holding cost $c_{\mathrm{hold}} = \$0.01$\,M for the attacker, and the ransom $R_1$  escalates to $\$112.5$\,M.

At \textbf{Passes~2--3}, after some internal coordination, the defender initiates the two-pass \textsc{Restore (SM)} procedure. Across these two passes they incur $c_{\downarrow} = \$6$\,M, plus the safe mode cost of $C_j= \$0.01$\,M. Therefore the defender’s expected costs reaches $\$9.01$\,M, while the ransom $R_3$ has escalated to $\$124.5$\,M. For the attacker, three passes of holding cost have accrued, leaving the attacker with a negative $\$14.03$\,M payoff.

At \textbf{Pass~4}, Unfortunately, safe mode did not stop the ransomware attack. The equilibrium policy at pass~4 recommends the \textsc{Restore (PT)} strategy, but again, the defender lacks a procedure for issuing the privileged command. Thus, they are forced to remain in \textsc{Idle}. This pass adds another $c_{\downarrow} = \$3$\,M, raising the defender’s expected cost to $\$12.01$\,M, while the ransom reaches $R_4 = \$130.5$\,M. For the attacker, another holding cost of $\$0.01$\,M is incurred, for a $-14.04$\,M payoff.  

At \textbf{Pass~5}, the game has reached its terminal condition discussed in Sec.~\ref{subsec:backward-induction} which recommends paying the ransom as the least costly option. Therefore the U.S. Space Force is forced to pay the ransom amount $R_5 = \$136.5$\,M to \attacker and the game ends.
In total, the unprepared defender's expected cost is $\$148.51$\,M, combining four passes of downtime, the failed safe-mode attempt, and the final ransom payment. The attacker, after subtracting their exploit and holding costs, obtains a $\$122.46$\,M payoff.

\begin{table*}[h]
\centering
\caption{Per-pass accounting for the Prepared Defender realized scenario. All costs in millions of dollars.}
\small
\setlength{\tabcolsep}{6pt}
\renewcommand{\arraystretch}{1.05}
\begin{tabular}{@{}c c c c c c c c@{}}
\toprule
\makecell{\textbf{Pass}} &
\makecell{\textbf{Realized}\\\textbf{Strategy}} &
\makecell{\textbf{Ransom}} &
\makecell{\textbf{Defender}\\\textbf{Incremental}} &
\makecell{\textbf{Defender}\\\textbf{Cumulative}} &
\makecell{\textbf{Remaining}\\\textbf{Restores}} &
\makecell{\textbf{Attacker}\\\textbf{Incremental}} &
\makecell{\textbf{Attacker}\\\textbf{Cumulative}} \\
\midrule

1 & \textsc{Restore (SM)} & $\$112.5\,\text{M}$ &
\makecell[l]{$\$3\,\text{M}$ (downtime)} &
$\$3\,\text{M}$ & \{SM, PT\} &
\makecell[l]{$-\$14\,\text{M}$ (exploit)\\ $-\$0.01\,\text{M}$ (hold)} &
$-\$14.01\,\text{M}$ \\[4pt]

\midrule

2 & \textsc{Restore (SM)} & $\$118.5\,\text{M}$ &
\makecell[l]{$\$3.00\,\text{M}$ (downtime)\\ $\$0.01\,\text{M}$ (SM cost)} &
$\$6.01\,\text{M}$ & \{PT\} &
\makecell[l]{$-\$0.01\,\text{M}$ (hold)} &
$-\$14.02\,\text{M}$ \\[4pt]

\midrule

3 & \textsc{Restore (PT)} & $\$124.5\,\text{M}$ &
\makecell[l]{$\$3.00\,\text{M}$ (downtime)\\ $\$0.01\,\text{M}$ (PT cost)} &
$\$9.02\,\text{M}$ & \{\} &
\makecell[l]{$-\$0.01\,\text{M}$ (hold)} &
$-\$14.03\,\text{M}$ \\[4pt]

\midrule

4 & Game Ends & $\$130.5\,\text{M}$ &
\makecell[l]{$\$0.00\,\text{M}$} &
$\$9.02\,\text{M}$ & \{\} &
\makecell[l]{$\$0.00\,\text{M}$} &
$-\$14.03\,\text{M}$ \\[4pt]

\midrule
\rowcolor{gray!10}
\multicolumn{3}{r}{} &
\multicolumn{3}{c}{\textbf{Defender Expected Cost = \$9.02\,M}} &
\multicolumn{2}{c}{\textbf{Attacker Payoff = --\$14.03 M}} \\

\bottomrule
\end{tabular}
\label{tab:realized}
\end{table*}

\subsection{Prepared Defender}

We now describe the strategies that the \emph{prepared defender} implements. At each pass we compare the \emph{realized selected strategy} against the strategy recommended by the equilibrium under ideal circumstances as shown in Table~\ref{tab:pass-summary}.

At \textbf{Passes~1--2}, the equilibrium recommends the \textsc{Restore (SM)} strategy, i.e., safe mode. The U.S. Space Force follows its previously tested and documented standard operating procedures (SOPs) and immediately initiates it. Over these two passes, the defender incurs $c_{\downarrow} = \$6$\,M in downtime and the safe mode execution cost of $C_j = \$0.01$\,M, and the ransom escalates to $R_2 = \$118.5$\,M. For the attacker, two passes of holding cost accrue, for a cumulative payoff of $\$-14.02$\,M.

At \textbf{Pass~3}, the U.S. Space Force determines that safe mode did not stop the attack. The equilibrium recommends \textsc{Restore (SM)} again, but this strategy cannot be reexecuted. Therefore, the defender implements \textsc{Restore (PT)}. This pass incurs the restore cost $C_k = \$0.01$\,M and one pass of downtime $c_{\downarrow} = \$3$\,M, and the ransom escalates to $R_3 = \$124.5$\,M. The attacker's cumulative payoff is $\$-14.03$\,M.

At \textbf{Pass~4}, the privileged telecommand succeeds, restoring nominal operations and liberating the hostage \examplesat.

In total, the prepared defender's expected cost is $\$9.02$\,M, composed of three passes of downtime, the safe-mode cost, and the privileged-telecommand cost, while avoiding any ransom payment. The attacker, after subtracting the exploit and holding costs, ends with a negative payoff of $-14.03$\,M.

The realized prepared defender game decision tree is illustrated in Fig.~\ref{fig:tree}, with its per-pass economic parameters accounting summarized in Table~\ref{tab:realized}.

\subsection{Realized Scenarios Summary}

Under our model’s finite-pass structure, escalating ransom, and costly downtime, the unprepared defender loses critical early passes to \textsc{Idle}, triggering compounding delays resulting in a $\$148.51$\,M expected cost and an attacker payoff of $\$122.46$\,M. The prepared defender, by contrast, executes both restore actions within the model’s tight pass constraints, avoids any ransom payment, and liberates the \examplesat for only $\$9.02$\,M while leaving the attacker with a $\$-14.03$\,M payoff. These outcomes, show that mission operational preparation, is a key factor in satellite ransomware attacks.

\section{Discussion}

\subsection{Real-World Implications for Satellite Mission Operations}

\textbf{Operational Readiness.} Mission operators should document their available recovery strategies and test them in simulations, digital twins or testbeds so that their success rates, costs, and duration (number of required passes) are known.

\textbf{Increase Contact Opportunities.} Our model reveals that contact opportunities are crucial during an attack. As such, operators should increase the number of contact opportunities for each pass, e.g., increasing the number of ground stations. This could be accomplished by deploying mobile ground stations or creating partnerships with other missions.

\textbf{Response Planning and Training.} Operators should conduct wargames using the orbital escalation game to gain experience and prepare standard operating procedures in the event of a ransomware attack. This would allow operators to make rational decisions under pressure and avoid mistakes.

These implications may resemble conventional mission practices, but their security motivation, pass-dependent urgency, and quantified economic justification are new.

\subsection{Limitations}

\textbf{Modeled Orbital Regimes.} Our model relies on orbital passes to dictate the phase of the game, making our model applicable to LEO and MEO satellites only as GEO satellites are always in communication with their ground stations.

\textbf{Perfect Information Assumption.} Under our model's perfect information assumption, the attacker is assumed to know the defender's set of recovery procedures and may therefore plan actions to mitigate them. In practice, however, successful "counter-recovery" strategies  require additional attacker capabilities (e.g., uplink jamming) and are out of the scope of our model.
Additionally, in practice, an attacker may not always know which specific recovery procedure the defender has attempted.
In general, our Stackelberg model's perfect information assumption provides a conservative worst-case baseline that ensures that the defender’s equilibrium strategies remain robust even under maximal attacker awareness.

\textbf{Rationality Assumption.} Our model assumes that all players are fully rational and assumes that each player seeks to maximize expected utility given the available information and constraints. In practice, real-world attackers may exhibit bounded rationality, pursue non-economic objectives, or deviate from equilibrium behavior due to political reasons.

\textbf{Discrete Passes.}
The current model uses discrete orbital passes as contact opportunities in satellite operations. In practice, however, real-world missions may involve multiple geographically distributed ground stations, inter-satellite links (ISLs), and heterogeneous uplink capabilities, allowing both defenders and attackers to interact with a spacecraft outside of a single line-of-sight pass. These factors can increase the number, duration, and overlap of effective interaction opportunities, and thus are not fully captured by a single-pass abstraction. 

\subsection{Generalizability}
The orbital escalation game provides a general analytical framework for studying attacker-defender interactions in space systems that unfold over repeated, constrained control opportunities. While this work instantiates the model in the context of ransomware, the same game structure can be applied to other classes of space security scenarios.

For example, coordinated jamming~\cite{let2024planta} campaigns involving multiple rogue ground stations targeting subsets of a LEO constellation can be modeled as interaction cycles, where both attacker interference allocation and defender mitigation decisions are shaped by geography-dependent access and cost.

\section{Related Work}

Prior game-theoretic models of ransomware primarily focus on enterprise or generic network environments. 
\emph{Laszka et~al.}~\cite{laszka2017economics} developed the first formal ransomware model as a two-stage game between organizations and an attacker, emphasizing backup investments and ransom decisions. 
Their model captures key economic incentives but remains static, with limited strategic depth and no temporal evolution. 
\emph{Caporusso et~al.}~\cite{caporusso2019game} later proposed a descriptive finite-state machine model capturing sequential escalation and negotiation but without economic parameters. 
\emph{Li and Liao}~\cite{li2021game} proposed a model that includes enterprise settings, integrating risk perception and insurance but retaining a single-stage structure.  

In contrast, our model introduces a multi-stage ransomware game tailored to satellite operations. 
Our model expands the defender's action space to four strategies, incorporates seven economic parameters derived from mission costs and recovery outcomes, and unfolds over multiple orbital passes. 
As depicted in Table~\ref{tab:comparison_numbers}, our model offers the most comprehensive formulation of ransomware dynamics to date, integrating richer economic factors and defender strategies.

\begin{table}[t]
\centering
\caption{Comparison between our work and previous work}
\setlength{\tabcolsep}{3.5pt} 
\renewcommand{\arraystretch}{0.9} 
\begin{tabular}{lcccc}
\toprule
\textbf{Work} &
\makecell{\textbf{Defender}\\\textbf{Strategies}} &
\makecell{\textbf{Economic}\\\textbf{Params}} &
\makecell{\textbf{Game}\\\textbf{Stages}} &
\textbf{Domain} \\
\midrule
Laszka et al.\,(2017)~\cite{laszka2017economics} & 2 & 7 & 2 & Terrestrial \\
Caporusso et al.\,(2019)~\cite{caporusso2019game} & 3 & 5 & 3 & Terrestrial \\
Li et al. (2021)~\cite{li2021game} & 2 & 6 & 4 & Terrestrial \\
\textbf{Orbital Escalation Game} & \textbf{4} & \textbf{7} &\makecell{\textbf{Multiple}\\\textbf{Passes}} & \textbf{Space} \\
\bottomrule
\end{tabular}
\label{tab:comparison_numbers}
\end{table}

\section{Future Work}

The orbital escalation game model opens several directions for extension. First, we plan to relax the perfect information assumption by formulating a Bayesian game~\cite{gibbons1997introduction} extension that captures uncertainty and partial knowledge between attacker and defender.

Additionally, future work may refine the abstraction of communication windows by parameterizing them using deployment-specific characteristics, including the number and geographic distribution of ground stations, contact durations measured in seconds or minutes, and uplink redundancy for both defenders and attackers. 

Beyond orbital passes, the framework can be extended to capture alternative ransomware scenarios. For example, launch segment attacks, reentry or crewed spacecraft scenarios which may be better modeled as one-shot rather than sequential games. Related game theory models have been explored in other space settings, such as docking operations~\cite{hiramatsu2007game}.

Another direction is a sensitivity analysis~\cite{gibbons1997introduction} of the model to characterize how equilibrium strategies change as key parameters vary. By exploring mission loss, or ransom, such analyses can identify how and why equilibria shifts and derive new satellite mission operation recommendations.

Finally, an important extension is to move from single asset analysis to constellation-level interactions. In such settings, attackers may selectively disrupt subsets of satellites, while defenders must consider partial service degradation, recovery prioritization, and alterative recovery vectors such as ISLs.

\section{Conclusion}

Although no satellite ransomware incident has yet been documented, research shows that they are feasible. 
Satellite operators and policymakers should therefore be prepared.
In this paper, we introduced the \emph{orbital escalation game}, the first game-theoretic model for satellite ransomware attacks. We hope that satellite owners, and researchers employ our model, and its open-source analysis tool, to plan their responses before, during, and after a spacecraft is held for ransom.

\bibliographystyle{IEEEtran}
\bibliography{bib}

@article{gibbons1997introduction,
  title={An introduction to applicable game theory},
  author={Gibbons, Robert},
  journal={Journal of Economic Perspectives},
  volume={11},
  number={1},
  pages={127--149},
  year={1997},
  publisher={American Economic Association}
}

@online{darkreading2022_apttools,
  title        = {APT Groups Make Quadruple What They Spend on Attack Tools},
  author       = {Dark Reading Staff},
  year         = {2022},
  url          = {https://www.darkreading.com/cyberattacks-data-breaches/apt-groups-make-quadruple-what-they-spend-on-attack-tools},
  note         = {Accessed: 2025-10-17},
  organization = {Dark Reading},
}

@misc{nasa_new_horizons_safe_mode_2017,
  title        = {New Horizons Exits Brief Safe Mode, Recovery Operations Continue},
  author       = {{National Aeronautics and Space Administration (NASA)}},
  year         = {2017},
  howpublished = {\url{https://www.nasa.gov/general/new-horizons-exits-brief-safe-mode-recovery-operations-continue/}},
  note         = {Accessed: YYYY-MM-DD}
}

@article{samir2025attitude,
  title={Attitude determination and control of gps satellites: Stabilization, orbital insertion, and operational control mechanisms},
  author={Samir, Oliullah},
  journal={arXiv preprint arXiv:2508.01660},
  year={2025}
}

@INPROCEEDINGS{let2024planta,
  author={Planta, Ulysse and Rederlechner, Julian and Marra, Gabriele and Abbasi, Ali},
  booktitle={2024 Security for Space Systems (3S)}, 
  title={Let Me Do It For You: On the Feasibility of Inter-Satellite Friendly Jamming}, 
  year={2024},
  volume={},
  number={},
  pages={1-6},
  doi={10.23919/3S60530.2024.10592285}}

@misc{NASA2022VoyagerPatch,
  author       = {{NASA Jet Propulsion Laboratory}},
  title        = {NASA’s Voyager Team Focuses on Software Patch for Thrusters},
  howpublished = {\url{https://www.nasa.gov/missions/voyager-program/nasas-voyager-team-focuses-on-software-patch-thrusters/}},
  year         = {2022},
  note         = {Accessed: 2025-10-15},
  organization = {NASA},
}

@inproceedings{KhojastehDana2023Logical,
  title        = {Logical Maneuvers: Detecting and Mitigating Adversarial Hardware Faults in Space},
  author       = {Fatemeh Khojasteh Dana and Saleh Khalaj Monfared and Shahin Tajik},
  booktitle    = {Network and Distributed System Security Symposium (NDSS)},
  year         = {2023},
  url          = {https://www.ndss-symposium.org/ndss-paper/auto-draft-622/},
}

@misc{federalnews2011_stuxnet,
  title        = {Stuxnet holds appeal as alternative weapon of war},
  author       = {{Federal News Network}},
  year         = {2011},
  month        = sep,
  howpublished = {\url{https://federalnewsnetwork.com/the-federal-drive-with-terry-gerton/2011/09/stuxnet-holds-appeal-as-alternative-weapon-of-war/\#:~:text=He\%20deconstructed\%20the%20Stuxnet\%20source,And\%20nobody\%20died}},
  note         = {Accessed: 2025-10-02}
}

@misc{gomspaceA3200,
  author       = {{GomSpace}},
  title        = {NanoMind A3200 On-board Computer},
  year         = {2023},
  howpublished = {\url{https://gomspace.com/shop/subsystems/on-board-computers/nanomind-a3200.aspx}},
  note         = {Accessed: 2025-09-19}
}

@inproceedings{willbold2023space,
  title={Space odyssey: An experimental software security analysis of satellites},
  author={Willbold, Johannes and Schloegel, Moritz and V{\"o}gele, Manuel and Gerhardt, Maximilian and Holz, Thorsten and Abbasi, Ali},
  booktitle={2023 IEEE Symposium on Security and Privacy (SP)},
  pages={1--19},
  year={2023},
  organization={IEEE}
}

@misc{eoportalGioveA,
  author       = {{ESA / eoPortal Directory}},
  title        = {GIOVE-A (Galileo In-Orbit Validation Element-A) Satellite Mission},
  year         = {2024},
  howpublished = {\url{https://www.eoportal.org/satellite-missions/giove-a}},
  note         = {Accessed: 2025-09-19}
}

@misc{beesatRecovery,
  author       = {PistonMiner},
  title        = {How a German researcher resurrected BEESAT-1 after more than a decade},
  year         = {2024},
  howpublished = {\url{https://en.clickpetroleoegas.com.br/como-um-hacker-alemao-ressuscitou-um-satelite-abandonado-ha-12-anos-e-transformou-o-impossivel-em-realidade/}},
  note         = {Accessed: 2025-09-19}
}

@inproceedings{laszka2017economics,
  title={On the economics of ransomware},
  author={Laszka, Aron and Farhang, Sadegh and Grossklags, Jens},
  booktitle={International conference on decision and game theory for security},
  pages={397--417},
  year={2017},
  organization={Springer}
}

@article{o2019economic,
  title={Economic benefits of the global positioning system (GPS)},
  author={O'Connor, Alan C and Gallaher, Michael P and Clark-Sutton, Kyle and Lapidus, Daniel and Oliver, Zack T and Scott, Troy J and Wood, Dallas W and Gonzalez, Manuel A and Brown, Elizabeth G and Fletcher, Joshua},
  year={2019},
  publisher={RTI International Research Triangle Park, NC}
}

@online{spacenews_falcon9_gps3,
  author    = {Foust, Jeff},
  title     = {SpaceX wins \$82 million contract for 2018 Falcon 9 launch of GPS 3 satellite},
  year      = {2016},
  url       = {https://spacenews.com/spacex-wins-82-million-contract-for-2018-falcon-9-launch-of-gps-3-satellite/},
  note      = {Accessed: 2025-09-15},
  publisher = {SpaceNews}
}

@online{spacenews_gps744,
  author    = {Erwin, Sandra},
  title     = {Space Force orders three GPS satellites for \$744 million},
  year      = {2023},
  url       = {https://spacenews.com/space-force-orders-three-gps-satellites-for-744-million/},
  note      = {Accessed: 2025-09-15},
  publisher = {SpaceNews}
}

@book{osborne2004introduction,
  title={An introduction to game theory},
  author={Osborne, Martin J and others},
  volume={3},
  number={3},
  year={2004},
  publisher={Springer}
}

@book{bellman2015applied,
  title={Applied dynamic programming},
  author={Bellman, Richard E and Dreyfus, Stuart E},
  year={2015},
  publisher={Princeton university press}
}

@inproceedings{liu2025teamwork,
author = {Liu, Siyu and Bazzi, Rida and Fang, Fei and Bao, Tiffany},
title = {Teamwork Makes the Defense Work: Comprehensive Vulnerability Defense Resource Allocation},
year = {2025},
isbn = {9798400714269},
publisher = {International Foundation for Autonomous Agents and Multiagent Systems},
address = {Richland, SC},
booktitle = {Proceedings of the 24th International Conference on Autonomous Agents and Multiagent Systems},
pages = {1362–1370},
numpages = {9},
location = {Detroit, MI, USA},
series = {AAMAS '25}
}

@misc{sophos2024ransomware,
  author       = {{Sophos}},
  title        = {Ransomware Payments Increase 500\% Last Year, Finds Sophos State of Ransomware 2024},
  year         = {2024},
  url          = {https://www.sophos.com/en-us/press/press-releases/2024/04/ransomware-payments-increase-500-last-year-finds-sophos-state},
  note         = {Accessed: 2025-07-16},
  institution  = {Sophos},
}

@article{li2021game,
  title={Game theory of data-selling ransomware},
  author={Li, Zhen and Liao, Qi},
  journal={Journal of Cyber Security and Mobility},
  pages={65--96},
  year={2021}
}

@article{reese2025space,
  title   = {Space assets could be held ransom. Will we have any choice but to pay?},
  author  = {Reese, Nick},
  journal = {SpaceNews},
  year    = {2025},
  month   = {June},
  url     = {https://spacenews.com/space-assets-could-be-held-ransom-will-we-have-any-choice-but-to-pay/},
  note    = {Accessed: 2025-06-03}
}

@inproceedings{hiramatsu2007game,
  title={Game theoretic approach to post-docked satellite control},
  author={Hiramatsu, Takashi and Fitz-Coy, Norman G},
  booktitle={Proceedings of the 20th International Symposium on Space Flight Dynamics},
  year={2007}
}

@inproceedings{rederlechner2026ota,
  title     = {One Small Patch for a File, One Giant Leap for OTA Updates},
  author    = {Rederlechner, Julian and Planta, Ulysse and Abbasi, Ali},
  booktitle = {Proceedings of the Workshop on Security of Space and Satellite Systems (SpaceSec)},
  year      = {2026},
  address   = {San Diego, CA, USA},
  month     = feb,
}

@misc{viasat_attack,
  title={Case Study: Viasat Attack},
  howpublished={\url{https://cyberconflicts.cyberpeaceinstitute.org/law-and-policy/cases/viasat}},
  note={Accessed: 2025-05-24}
}

@article{donchev2024ransomware,
  title={Evaluating an Effective Ransomware Infection Vector in Low Earth Orbit Satellites},
  author={Donchev, Marin and Smyth, Dylan},
  journal={arXiv preprint arXiv:2412.04601},
  year={2024}
}

@misc{USSF_GPSIII_SVN77_2020,
  author       = {United States Space Force},
  title        = {SMC and its Partners Poised to Launch Fourth {GPS} {III} Satellite},
  howpublished = {\url{https://www.losangeles.spaceforce.mil/News/Article-Display/Article/2365822/smc-and-its-partners-poised-to-launch-fourth-gps-iii-satellite/}},
  note         = {Accessed: 2025-12-02},
  year         = {2020},
  month        = sep,
  publisher    = {U.S. Space Force Los Angeles Garrison, Public Affairs}
}

@inproceedings{caporusso2019game,
  title={A game-theoretical model of ransomware},
  author={Caporusso, Nicholas and Chea, Singhtararaksme and Abukhaled, Raied},
  booktitle={Advances in Human Factors in Cybersecurity: Proceedings of the AHFE 2018 International Conference on Human Factors in Cybersecurity, July 21-25, 2018, Loews Sapphire Falls Resort at Universal Studios, Orlando, Florida, USA 9},
  pages={69--78},
  year={2019},
  organization={Springer}
}

@INPROCEEDINGS{falco2023wannafly,
  author={Falco, Gregory and Thummala, Rajiv and Kubadia, Arpit},
  booktitle={2023 IEEE 9th International Conference on Space Mission Challenges for Information Technology (SMC-IT)}, 
  title={WannaFly: An Approach to Satellite Ransomware}, 
  year={2023},
  volume={},
  number={},
  pages={84-93},
  doi={10.1109/SMC-IT56444.2023.00018}}

@book{fudenberg1991game,
  title={Game Theory},
  author={Fudenberg, Drew and Tirole, Jean},
  year={1991},
  publisher={MIT Press}
}

@book{stackelberg1934marktform,
  title={Marktform und Gleichgewicht},
  author={von Stackelberg, Heinrich},
  year={1934},
  publisher={Springer},
  note={English translation: Market Structure and Equilibrium, 1952}
}

@article{nash1950equilibrium,
  title={Equilibrium points in n-person games},
  author={Nash, John F},
  journal={Proceedings of the National Academy of Sciences},
  volume={36},
  number={1},
  pages={48--49},
  year={1950},
  publisher={National Acad Sciences}
}

@book{von1944theory,
  title={Theory of Games and Economic Behavior},
  author={Von Neumann, John and Morgenstern, Oskar},
  year={1944},
  publisher={Princeton University Press}
}

\appendices

\section{\examplesat Case Study Parameter Details}\label{appendix:case-study-parameters}

\textbf{Mission Loss.} For mission loss we added the estimated cost for procuring the satellite and the cost for launching it. 
In 2022 it was reported that two new GPS III satellites had been ordered for \$744M~\cite{spacenews_gps744}. Using this figure we can infer that a single satellite costs around \$372M.
Additionally, in 2016 it was reported that the cost of launching a GPS III satellite was \$82M~\cite{spacenews_falcon9_gps3}. Adding these makes up for a total of \$454M which we round to \$450M.

\textbf{Downtime cost.} According to a 2019 report~\cite{o2019economic}, the U.S. economy derived approximately \$68.7 billion in benefits from GPS in 2017 alone. Assuming an active constellation of 31 satellites~\cite{USSF_GPSIII_SVN77_2020}, this implies an average contribution of \$2.2 billion per satellite per year. Dividing this evenly over time, we estimate a per-satellite downtime cost of approximately \$6M per day, or \$3M per orbital pass (assuming two passes per day). We adopt a \$3M per pass downtime cost.

\textbf{Restore Strategies.} Satellites, including the GPS III satellites have a safe mode~\cite{samir2025attitude} that allows them to shutdown non-essential systems. Additionally, engaging safe mode may take several hours or days~\cite{nasa_new_horizons_safe_mode_2017} which is why we decided to assign the safe mode strategy a restore duration of 2 passes. Because safe mode is a standard, contingency procedure routinely tested during spacecraft operations, we assign it a relatively high probability of success (0.9).  

Regarding the Privileged TC strategy we could not find any evidence for its use in GPS III satellites, however, these types of telecommands have been documented in other satellites~\cite{willbold2023space} and we assigned a restore duration of 1 pass as it involves sending one telecommand. Due to its less standardized use we assign a lower success probability (0.4).

\textbf{Initial Ransom.} Following the strategic attacker framework by Laszka et al.~\cite{laszka2017economics}, we set the ransom below the defender’s refusal loss to maximize the likelihood of payment while ensuring attacker profit. We conservatively set the ransom at 25\% or \$112.5M of the estimated mission loss (\$450M).

\textbf{Ransom Escalation.} To ensure that delaying payment is economically unattractive, the per-pass ransom escalation 
\(\Delta R\) should be set larger than the defender’s per-pass downtime cost \(c_{\mathrm{down}}\). As a conservative baseline, we set the escalation to twice the downtime cost: \$6M.

\textbf{One-time Exploit.} To estimate the cost of the one-time exploit we looked at previous instances of the cost of developing malware for Cyber-Physical Systems, specifically, Stuxnet. It is estimated that Stuxnet’s development costed \$10M in 2010~\cite{federalnews2011_stuxnet}, which equals \$14M in 2025 dollars.

\textbf{Holding Cost.} We estimate the attacker’s per-pass holding cost based on the recurring expenses of sustaining APT infrastructure. Reports indicate that APT groups invest up to \$10{,}000 in tools for their activities~\cite{darkreading2022_apttools}. As such we assign a conservative holding cost of \$10{,}000 per pass.

These parameter descriptions strengthen the realism of the fictional attack described in Sec.~\ref{sec:worked-example-once}.

\section{Model Parameter Summary}\label{appendix:parameterstable}

Table~\ref{tab:variables_summary} summarizes all of our model's parameters (Sec.~\ref{sec:model}) and provides information on how each parameter affects the game's equilibrium and its real-world satellite operational relevance.

\begin{table}[h]
\centering
\caption{Summary of Model Parameters and Symbols}
\label{tab:variables_summary}
\small
\setlength{\tabcolsep}{4pt}
\renewcommand{\arraystretch}{1.05}
\begin{tabular}{@{}p{3.1cm} p{1.4cm} p{4cm}@{}}
\toprule
\multicolumn{1}{c}{\textbf{Parameter}} &
\multicolumn{1}{c}{\textbf{Symbol}} &
\multicolumn{1}{c}{\textbf{Effect on Equilibrium}} \\
\midrule

Initial ransom demand &
$R_0$ &
Higher initial ransom accelerates Pay/Refuse threshold; extreme demands trigger immediate Refuse. \\

\midrule
Ransom escalation per pass &
$\Delta R$ &
Faster escalation pushes the defender toward early Pay/Refuse; very steep escalation makes Refuse dominant. \\

\midrule
Downtime cost per pass &
$c_{\downarrow}$ &
High $c_{\downarrow}$ pushes defender toward Pay; low $c_{\downarrow}$ enables Idle/Restore strategies. \\

\midrule
Restore cost (strategy $j$) &
$C_j$ &
High $C_j$ makes restore strategies unattractive or infeasible. \\

\midrule
Restore duration (in passes) &
$d_j$ &
Long $d_j$ makes restore unattractive when $c_{\downarrow}$ is high or when few passes remain. \\

\midrule
Restore success probability &
$p_j$ &
High $p_j$ makes Restore optimal; low $p_j$ favors Pay or Idle. \\

\midrule
Mission replacement cost / terminal loss &
$L_{\mathrm{ref}}$ &
If $L_{\mathrm{ref}} < R_k$, defender Refuses; if $L_{\mathrm{ref}} > R_k$, Pay may dominate. \\

\midrule
Attacker exploit cost &
$C_{\mathrm{atk}}$ &
High attacker cost discourages launching or escalating attacks. \\

\midrule
Attacker holding cost per pass &
$c_{\mathrm{hold}}$ &
High $c_{\mathrm{hold}}$ penalizes long games; benefits the defender when choosing Idle. \\

\midrule
Game horizon (number of passes) &
$K$ &
Short horizon forces early Pay/Refuse; long horizon supports Idle/Restore. \\
\bottomrule
\end{tabular}
\end{table}

\section{Ransomware Comparison}\label{appendix:parameterstable}
Table~\ref{tab:ransomware_comparison} showcases the differences between terrestrial and satellite ransomware attacks which are further discussed in Sec.~\ref{background:ransomwareinsatellites}.

\begin{table}[h]
\centering
\caption{Comparison of Ransomware in Terrestrial vs. Space Systems}
\label{tab:ransomware_comparison}
\small
\setlength{\tabcolsep}{2pt} 
\renewcommand{\arraystretch}{1.0}
\begin{tabular}{@{}>{\raggedright\centering\arraybackslash}p{1.6cm}
                >{\centering\arraybackslash}p{3.4cm}
                >{\centering\arraybackslash}p{3.4cm}@{}}
\toprule
\textbf{Aspect} &
\makecell{\textbf{Terrestrial}} &
\makecell{\textbf{Satellites}} \\
\midrule
\textbf{Execution Environment} &
\makecell[l]{• Windows/Linux\\• High CPU/RAM\\• Crypto libraries} &
\makecell[l]{• RTOS / bare-metal\\• Limited CPU/RAM\\• Often no crypto libraries} \\
\midrule
\textbf{Attack Vector} &
\makecell[l]{• Phishing\\• Leaked credentials} &
\makecell[l]{• Uplink entry\\• GS compromise} \\
\midrule
\textbf{Denial Mechanism} &
\makecell[l]{• Encrypt filesystem\\• Key-tied ransom} &
\makecell[l]{• Modify Flight Software\\• Disable FS exploit~\cite{falco2023wannafly}} \\
\midrule
\textbf{Impact} &
\makecell[l]{• Financial downtime\\• Local/regional} &
\makecell[l]{• Valuable asset loss~\cite{spacenews_gps744}\\• Nationwide/Global~\cite{o2019economic}} \\
\midrule
\textbf{Detection} &
\makecell[l]{• Network monitoring} &
\makecell[l]{• Telemetry anomalies} \\
\midrule
\textbf{Recovery} &
\makecell[l]{• Restore backups\\• Reimage systems\\• Cloud failover} &
\makecell[l]{• Firmware patch~\cite{NASA2022VoyagerPatch}\\• Safe mode~\cite{nasa_new_horizons_safe_mode_2017}\\• Privileged TC~\cite{willbold2023space}\\• Hardware reconfig~\cite{KhojastehDana2023Logical}\\ 
• OTA update~\cite{rederlechner2026ota}} \\ 
\bottomrule
\end{tabular}
\end{table}

\end{document}